\documentclass[12pt]{iopart}

\usepackage{graphicx}
\usepackage{subfigure}
\usepackage{color}
\usepackage[british]{babel}
\usepackage{color}
\usepackage{iopams}		
\usepackage[OT4]{fontenc}	
\usepackage{hyperref}
\usepackage{textcomp}
\usepackage{caption}

\newcommand{\bra}[1]{\langle #1|}
\newcommand{\ket}[1]{|#1\rangle}

\newcommand{\cre}{\hat{a}^{\dagger}}							

\newcommand{\bcre}{\hat{b}^{\dagger}}

\begin{document}

\title[Transverse Entanglement of Biphotons]{Transverse Entanglement of Biphotons}

\author{F Just$^1$, A Cavanna$^1$, M V Chekhova$^{1,2,3}$ and G Leuchs$^{1,3}$}

\address{$^1$ Max Planck Institute for the Science of Light, G\"unther-Scharowsky-Stra\ss{}e 1/Bau 24, 91058 Erlangen, Germany}

\address{$^2$ Department of Physics, M.V.Lomonosov Moscow State University, \\ Leninskie Gory, 119991 Moscow, Russia}

\address{$^3$ University of Erlangen-N\"urnberg, Staudtstrasse 7/B2, 91058 Erlangen, Germany}

\ead{felix.just@mpl.mpg.de}

\begin{abstract}
We measure the transverse entanglement of photon pairs on their propagation from the near to the far field of spontaneous parametric down-conversion (SPDC). The Fedorov ratio, depending on the widths of conditional and unconditional intensity measurements, is shown to be only able to characterise entanglement in the near and far field zones of the source. Therefore we also follow a different approach. By evaluating the first-order coherence of a subsystem of the state we can quantify its entanglement. Unlike previous measurements, which determine the Fedorov ratio via intensity correlations, our setup is sensitive to both phase and modulus of the biphoton state and thus always grants experimental access to the full transverse entanglement of the SPDC state. It is shown theoretically that this scheme represents a direct measurement of the Schmidt number.
\end{abstract}




\section{Introduction}
\label{Introduction}
Entanglement is an exciting phenomenon and a fundamental resource in quantum information and quantum computation. One convenient source of entanglement are photon pairs (biphotons) obtained by spontaneous parametric downconversion (SPDC). These photons can be entangled not only in discrete variables like polarisation or photon number but also position and momentum, which are continuous \cite{Law2004, Strekalov1995, Monken1998, Menzel2012}. The case of continuous variables is especially appealing for quantum informational tasks because it allows access to a larger Hilbert space \cite{Lloyd1999}. This is why entanglement in the transverse wavevectors of SPDC biphotons is currently in the focus of research \cite{Chiuri2012, Howell2012, Kang2012, Salakhutdinov2012, Braverman2012}.
Transverse entanglement can be understood in terms of the famous EPR scenario \cite{EPR}.
Consider a quantum state of two subsystems with the positions and momenta perfectly correlated. In such a system a measurement of position or momentum of one subsystem gives complete information about the corresponding variable in the other subsystem. The authors of \cite{EPR} argued that under the assumption of local reality, this is in disagreement with Heisenberg's uncertainty principle. As shown in \cite{Howell2004} the state of biphotons emitted by SPDC is an approximation of such an EPR state. In this context entanglement can be identified by violation of the inequality
\begin{equation}
\label{eqn:introduction:uncertainty}
\left( \Delta x \right)^2 \left( \Delta p \right)^2 \geq \left( \frac{\hbar}{2} \right)^2,
\end{equation}
where $\left( \Delta x \right)^2$ and $\left( \Delta p \right)^2$ are the variances of a quantum system in position and momentum respectively. 
The violation of equation (\ref{eqn:introduction:uncertainty}) has been measured \cite{Howell2004, Leach2012} and even though it does quantify the amount of transverse entanglement in the biphoton state, measurement in both near- (position) and far field (momentum) of the source are necessary to obtain a value. Closely related to the violation of (\ref{eqn:introduction:uncertainty}) is the Fedorov ratio \cite{Fedorov2004, Brida2009} which is especially appealing because it is an entanglement quantifier that can be directly measured. Unfortunately the Fedorov ratio does vary while the state propagates from the near to the far field region \cite{Chan2007} and even turns to unity at some point, indicating no entanglement. Thus the Fedorov ratio cannot be considered a measure for the full entanglement at any arbitrary position. In this work, we demonstrate a measurement of the full evolution of the Fedorov ratio between those two regimes. Additionally we implement a different scheme which has been proposed \cite{Chan2007} to fully quantify the transverse entanglement of the biphoton state. This measurement allows direct access to the Schmidt number \cite{Ekert1995,Law2004}, a quantity usually unattainable in the laboratory but always giving the full amount of entanglement.

\section{Fedorov Ratio}
\label{FedRatio}
\begin{figure}[<+htpb+>]
\begin{center}
\includegraphics[width=0.6\textwidth]{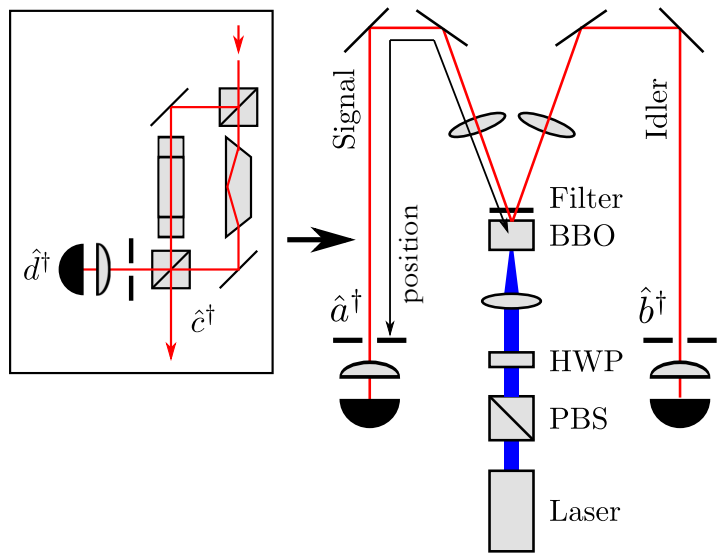}
\end{center}
\caption{Experimental setup to determine the Fedorov ratio. Measurements are performed at different $z$ positions by displacing the detectors. Inset on the left shows the modified Mach-Zehnder Interferometer we introduced in the signal arm, for the Schmidt number measurement.}
\label{fig:setup}
\end{figure}

\begin{figure}[<+htpb+>]
\subfigure{
\includegraphics[width=0.3\textwidth]{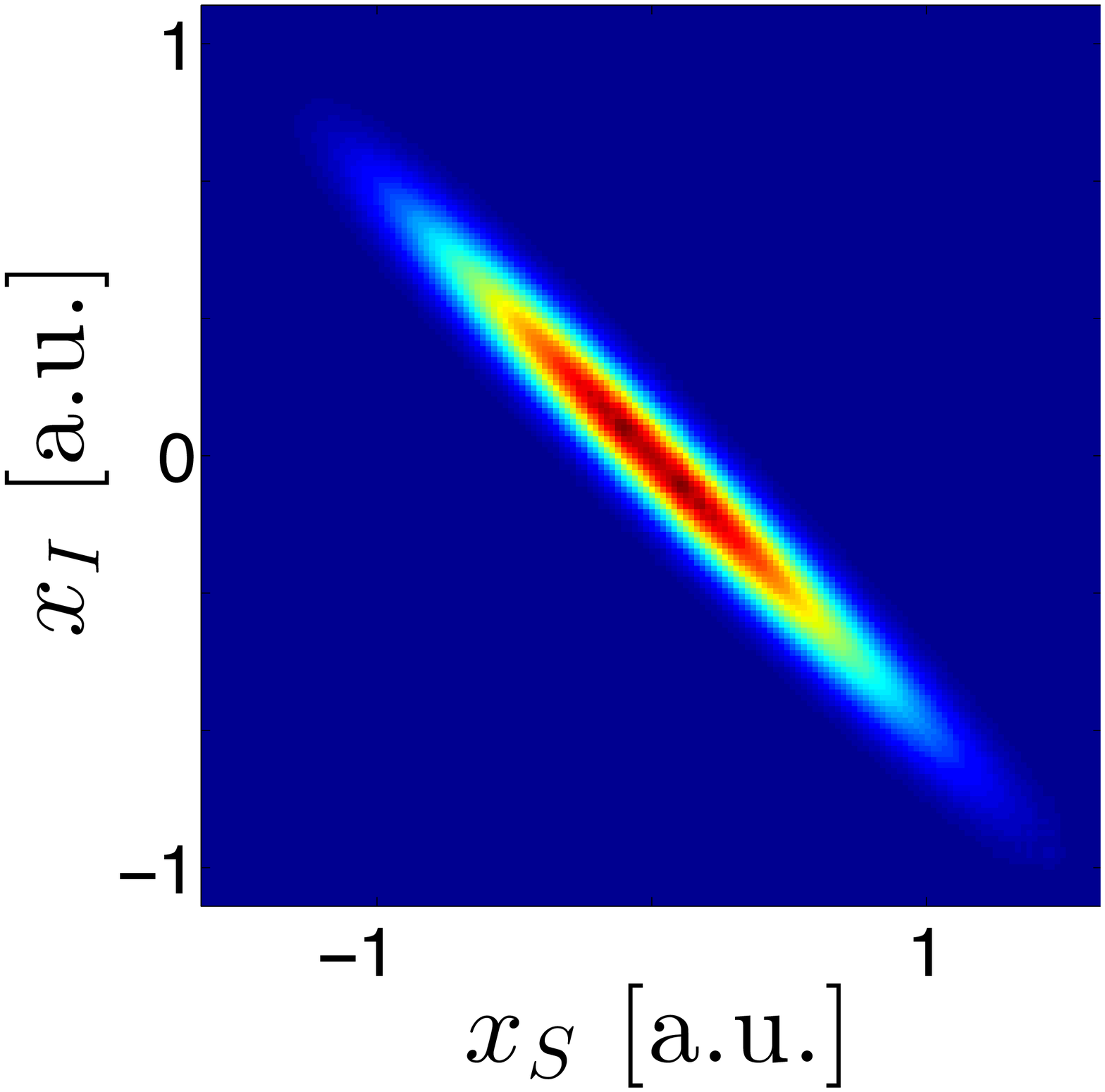}
}
\subfigure{
\includegraphics[width=0.3\textwidth]{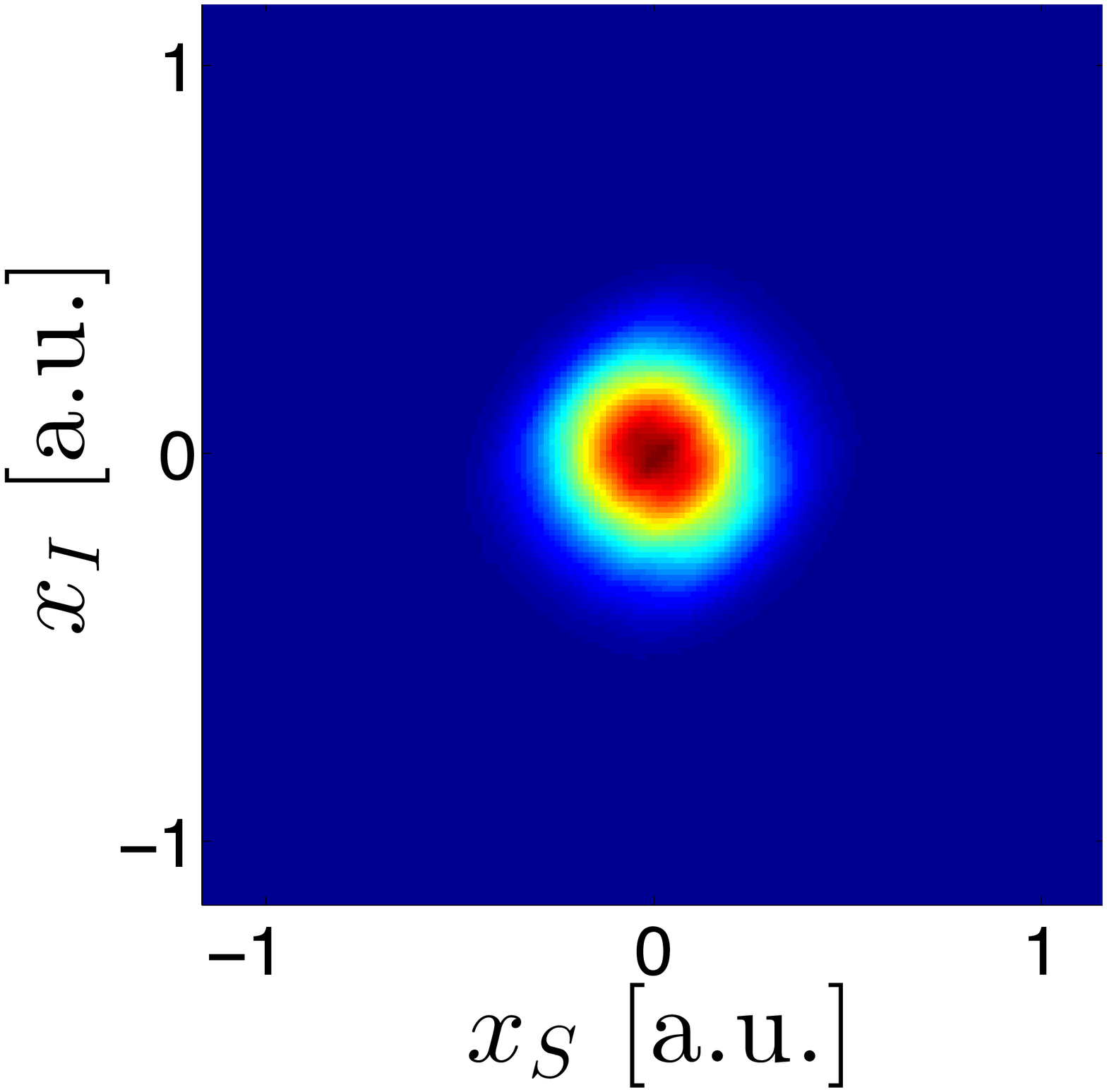}
}
\subfigure{
\includegraphics[width=0.3\textwidth]{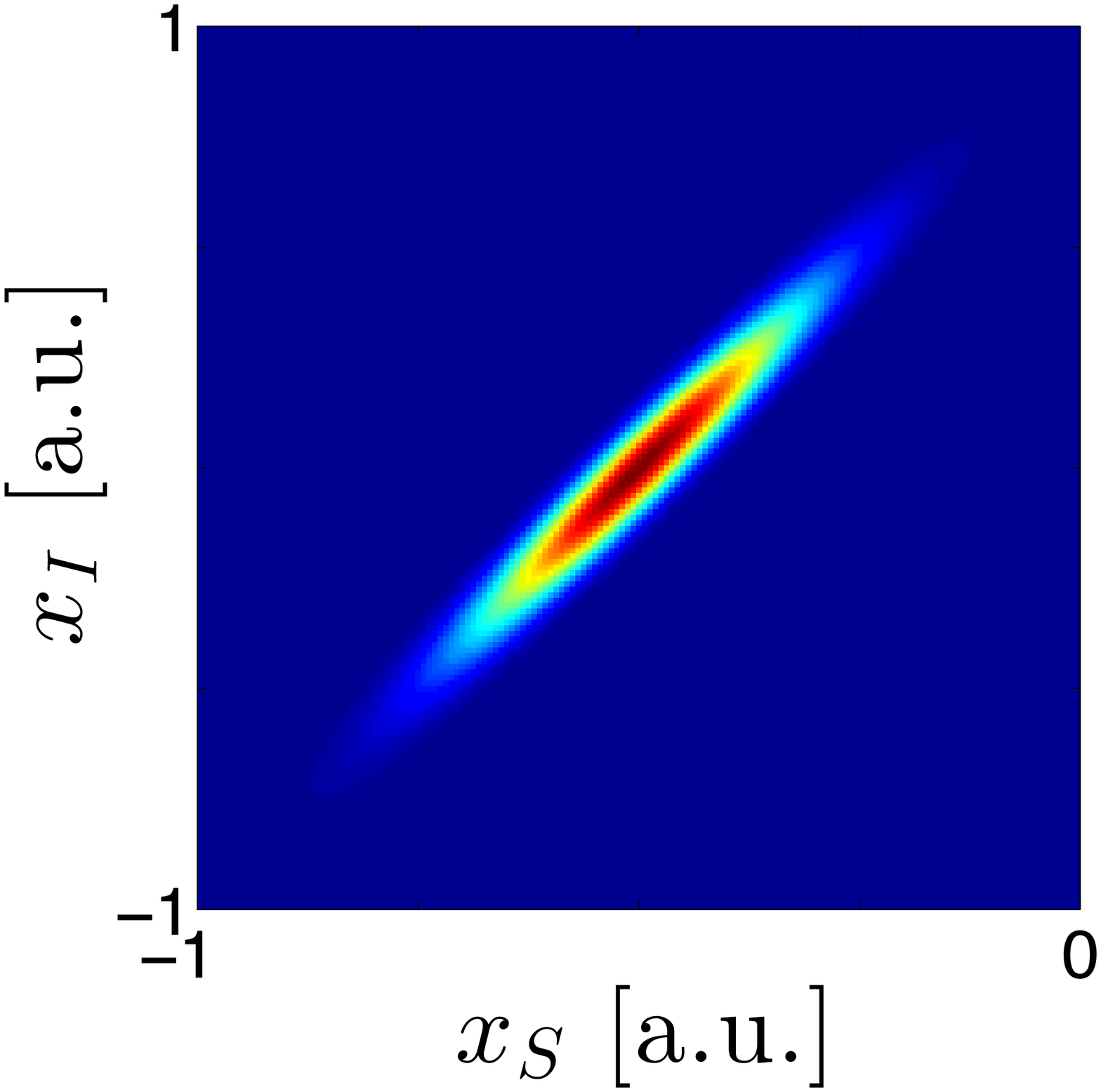}
}\\ \setcounter{subfigure}{0}
\subfigure[ Far field ($z=500$ mm)]{
\label{fig:marg:ff}
\includegraphics[width=0.32\textwidth]{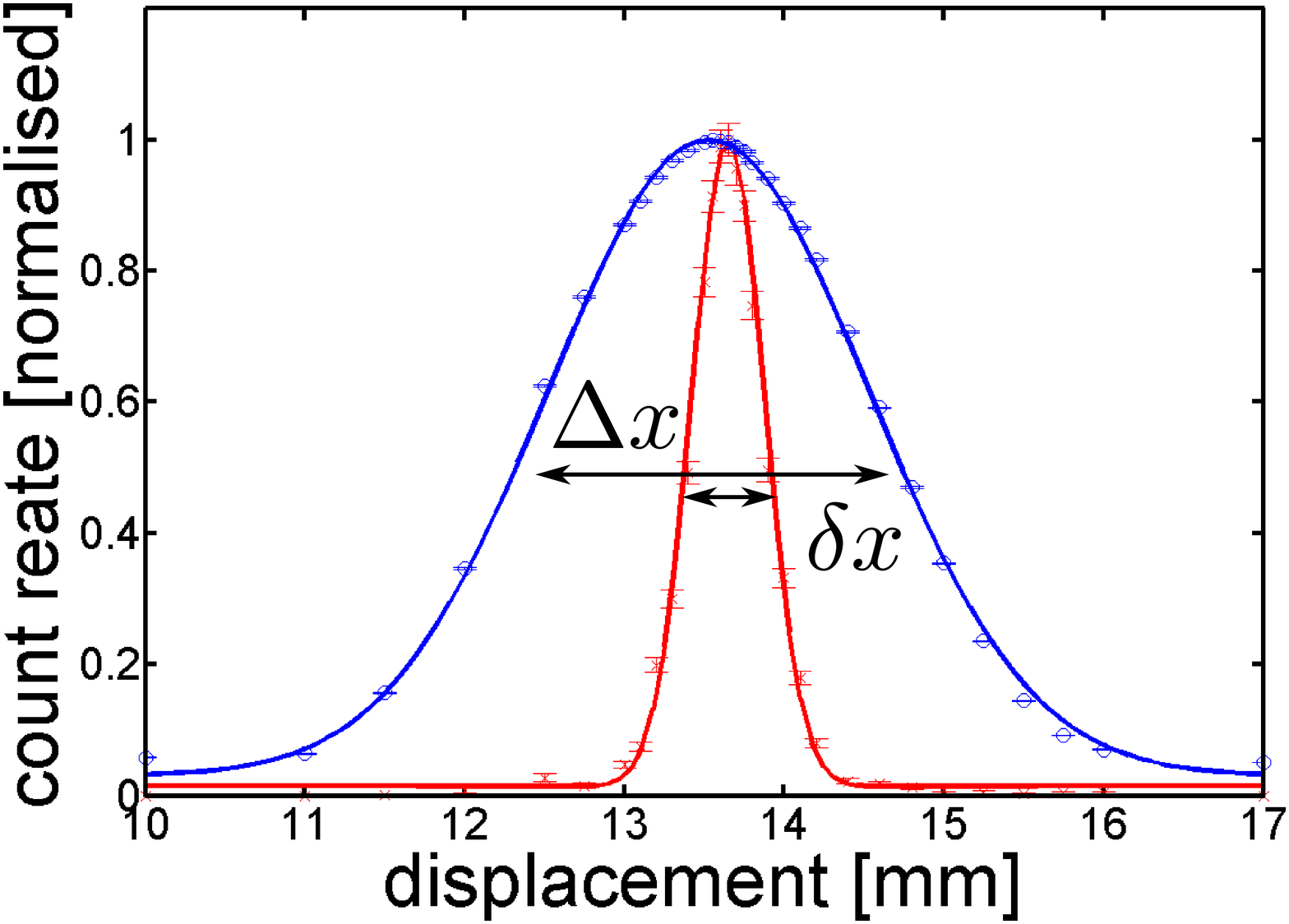}
}
\subfigure[ Intermedeate regime\newline($z=1440$ mm)]{
\label{fig:marg:if}
\includegraphics[width=0.32\textwidth]{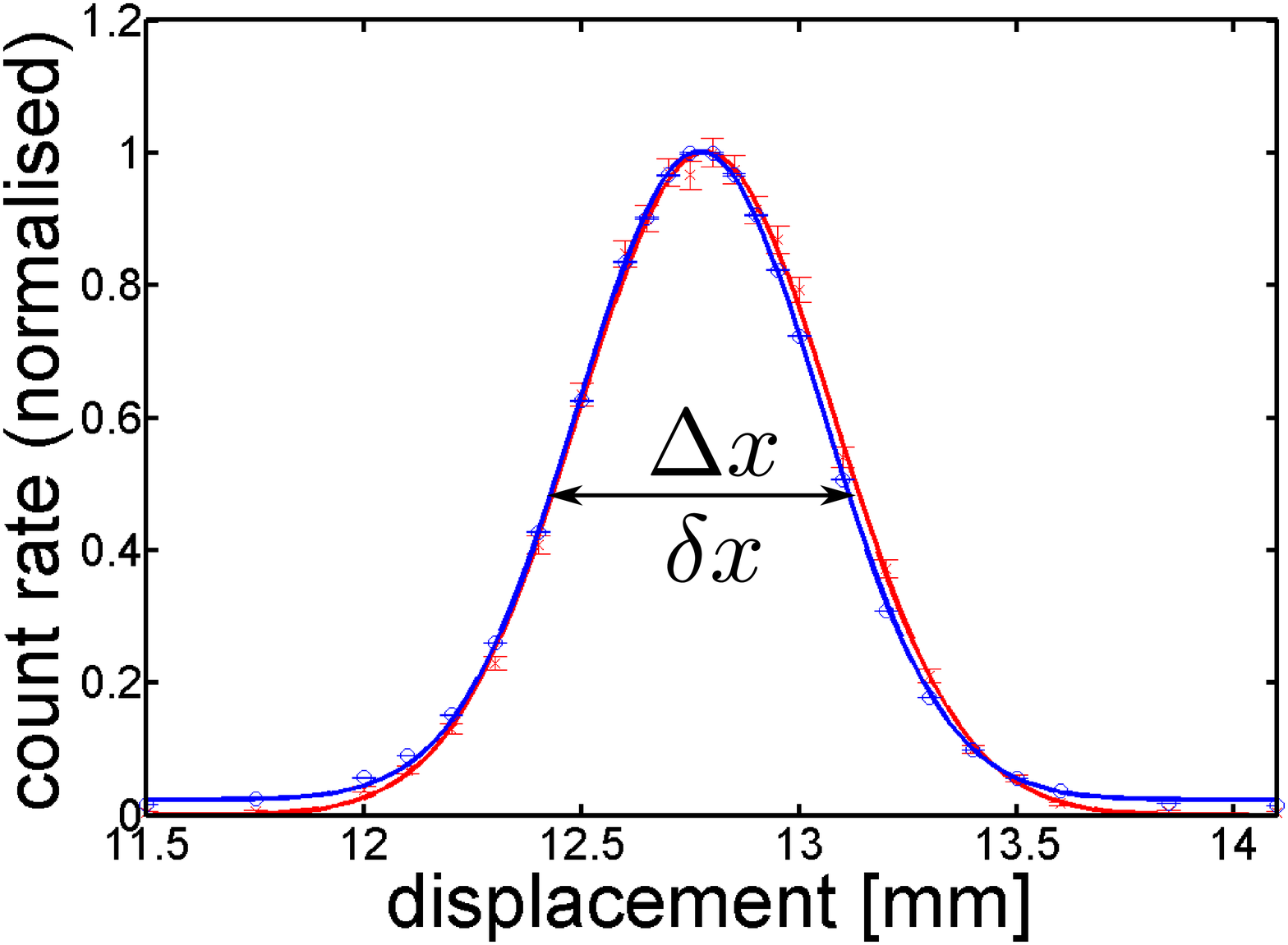}
}
\subfigure[ Near field  ($z=1550$ mm)]{
\label{fig:marg:nf}
\includegraphics[width=0.32\textwidth]{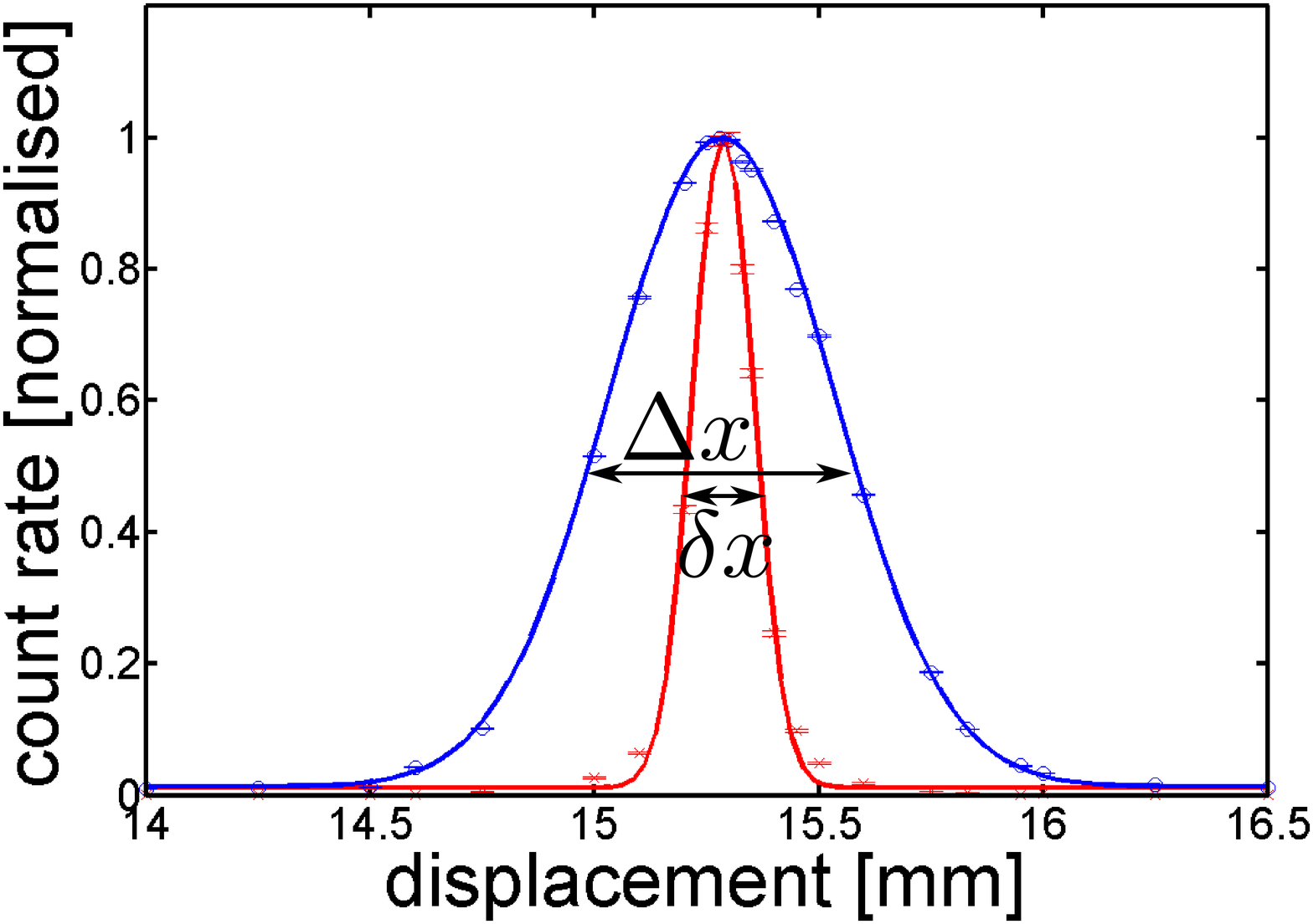}\\
}
\caption{Measured conditional (red) and unconditional (blue) photon distributions with simulated two-photon amplitudes, \ref{fig:marg:ff} in the far field (here the measured position $x$ corresponds to the actual momentum $p$), \ref{fig:marg:if} in the intermediate field where $R=1$, and \ref{fig:marg:nf} in the near field.}
\label{fig:fedMeasExamples}
\end{figure}

Consider the quantum state of a photon pair generated by SPDC at the distance $z$ from the centre of the crystal as \cite{Klyshko1988}
\begin{equation}
\label{eqn:FedRatio:PDCstate}
\ket{\psi \left( z \right)} = \int \int \mathrm{d} \vec{p} \, \mathrm{d} \vec{q} \, \Phi \left( \vec{p}, \vec{q}, z \right) \cre (\vec{p}) \, \bcre (\vec{q}) \ket{0}
\end{equation}
where $\vec{p}$ and $\vec{q}$ are the transverse wavevectors of the signal and idler photons and $\cre (\vec{p})$ and $\bcre (\vec{q})$ their respective creation operators. The properties of the state are governed by the two-photon amplitude $\Phi \left( \vec{p}, \vec{q}, z \right)$. The most widely used operational measure for the entanglement in such a system is the Fedorov ratio, which is given by 
\begin{equation}
\label{eqn:FedRatio:R}
R = \frac{\Delta p}{\delta p} = \frac{\Delta q}{\delta q}.
\end{equation}
Here $\Delta p$ is the standard deviation of the marginal angular distribution (or transverse wavevector spectrum) of the two photon-amplitude $P(\vec{p},z) = \int d \vec{q} \left| \Phi \left( \vec{p}, \vec{q}, z \right) \right|^2$, which we will from here on refer to as the unconditional distribution. The width $\delta p$ is given by the standard deviation of the conditional probability distribution $P(\vec{p}|\vec{q},z) = |\Phi \left( \vec{p}, \vec{q}, z \right)|^2$ at a fixed value of $\vec{q}$. Analogous expressions hold for $\vec{q}$. Another important feature of the two-photon amplitude is that both transverse dimensions are independent so that the two-photon amplitude factorises: $\Phi \left( \vec{p}, \vec{q} \right) = \Phi \left( p_x, q_x \right) \Phi \left( p_y, q_y \right)$. This allows us to study the behaviour of the system utilising only the spatial degree of freedom in $x$-direction.\\An intuitive approach to the Fedorov ratio is to think of entanglement as having two subsystems both of which are individually very uncertain (broad unconditional distributions) but at the same time exhibit very strong correlations between each other (narrow conditional distribution). In the EPR language this corresponds to having one particle whose momentum is almost impossible to predict but as soon as the second particle is measured the momentum of the first is precisely known.\\
The experimental setup is depicted on the right hand side of figure (\ref{fig:setup}). In order to generate biphotons we focus a cw laser with a measured beamwaist of of $245$\textmu m onto a $2$ mm BBO crystal. By means of an interference filter of $6$nm bandwidth and apertures positioned at the the correct emission angle, we select the degenerate non-collinear type I downconversion process: $404$nm $\rightarrow 808$nm. To select a certain part of the angular spectrum, we use slits of $30$\textmu m width and several mm height. The height of the slit leads to an integration over the $y$ direction in both conditional and unconditional distributions which does not change the ratio between them. Diffraction from the slits is compensated by cylindrical lenses before the light is fibrecoupled into avalanche photodiodes. For the measurement one of the slits remains fixed at the maximum intensity of the angular spectrum while the other one is scanned along transverse direction (in figure \ref{fig:setup} $x$, perpendicular to the propagation direction $z$) by means of a linear translation stage. We utilise a coincidence electronic circuit to record the coincidence-rates as well as single count rates of both detectors as a function of the transverse displacement. A pair of lenses ($f = 500$ mm) is employed to obtain the far field distribution of the angular photon spectra (which corresponds to the transverse momenta) in the focal plane. We apply Gaussian fits to both the conditional and the unconditional distributions and use their respective variances to calculate the Fedorov ratio via equation (\ref{eqn:FedRatio:R}).  A series of similar measurements is performed at various distances $z$ from the source up to the image plane of the lenses, where the near field (where the coordinate is equivalent to the position quadrature) of the photon distribution is found. The three measurements at positions of particular interest are shown in figure \ref{fig:fedMeasExamples}. In the same figure one also finds the numerical simulations of the two-photon amplitudes in the position representation. It is evident that there are correlations in the near-field zone and anti-correlations in the far field zone while no correlations are observed at a certain distance from the crystal.

\begin{figure}[<+htpb+>]
\begin{center}
\includegraphics[width=0.8\textwidth]{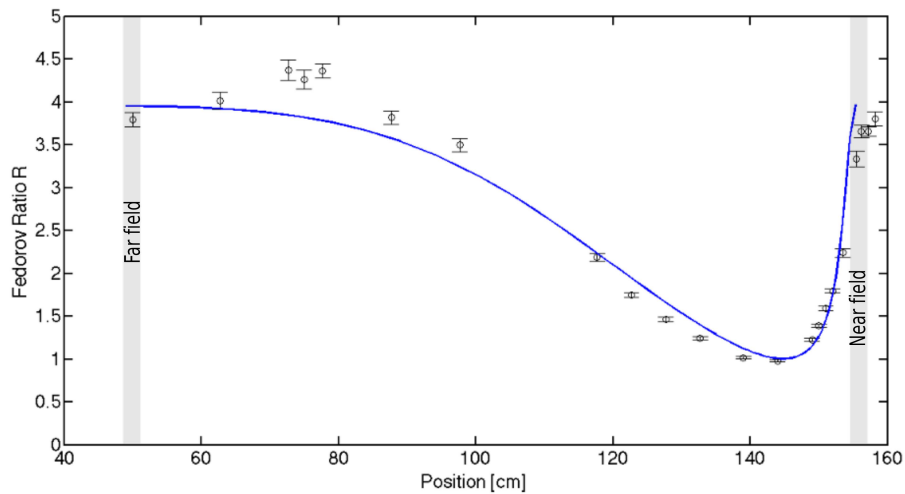}
\end{center}
\caption{Measurement of the Fedorov ratio from far- to near field: points show measurements, the solid line is a numerical simulation.}
\label{fig:Fedorov:resultsR}
\end{figure}

The dependence of the Fedorov ratio on the distance is plotted in figure \ref{fig:Fedorov:resultsR}. Note, that because of the effect of the lenses, the far field is obtained at a smaller distance (focal length) than the near field (image plane). It can be clearly seen how the Fedorov ratio varies as the state propagates through space. In the far field we obtain a Fedorov ratio $R = 4.0 \pm 0.3$. As we move our detectors towards the near field, the Fedorov ratio decreases. We even observe a drop to $R=1$ \cite{Chan2007} at a certain distance from the crystal, which would indicate no entanglement at this position before the complete entanglement emerges again in the near field. We additionally compare our measurement results with a numerical simulation (figure \ref{fig:Fedorov:resultsR}) and find the curve in good agreement with our measured data. An important issue with this kind of measure is that the Fedorov ratio is defined for the widths of Gaussian functions. In reality, the shape of the two-photon amplitude is better modelled by a Sinc function rather than a Gaussian and thus the concept of a width is not well defined in this case any more. The most curious feature of this experiment however is still the aforementioned decrease of the Fedorov ratio. Since we observe that the amount of entanglement increases again after the drop and even is completely restored in the near field, the entanglement cannot be lost due to decoherence.  So the question what happened to the `lost' entanglement remains. The explanation, given by Chan \textit{et. al.} \cite{Chan2007}, is that upon propagation, the entanglement of the state migrates from the wavefunction's modulus to its phase and back. Thus, since in the intermediate zone between near- and far field, the entanglement (or at least parts of it) resides in the phase of the state, it is inaccessible to intensity measurements such as the one we use to determine the Fedorov ratio. This prediction is clearly confirmed by our measurements.
As shown by Tasca \textit{et. al.} \cite{Tasca2008, Tasca2009}, application of different fractional Fourier transforms in signal and idler channels allows one to reveal intensity correlations in the intermediate zone. This is similar to the generalised quadrature measurements performed in \cite{Sych2009, Leuchs2009} and requires specific lens configurations for every distance $z$ in both signal and idler channels. In the following we present a more general approach, which works at arbitrary distances in the same configuration.\\ 

\section{Interferometric scheme and measurement}

Following a proposal by Chan \textit{et. al.} \cite{Chan2007}, we introduced a Mach-Zehnder-like interferometer in the signal channel. A sketch can be found in the inset on the left of figure \ref{fig:setup}. The interferometer contains one dove prism in each arm, one of which is rotated by $\pi / 2$ with respect to the other. Furthermore the interferometer had to be balanced up to the order of a few wavelengths. Due to internal reflection within the prisms, the beam in one arm of the interferometer is `spatially mirrored' in one direction while the beam in the other arm is `mirrored' orthogonally compared to the first. As this is equivalent to the inversion operation, one can say the beam is overlapped with its own inverted copy on the final beamsplitter. The output mode $\hat{c}^{\dagger}$ denotes the constructive output port of the interferometer while $\hat{d}^{\dagger}$ represents the destructive output mode. 
This kind of setup allows one to infer the degree of entanglement of the joint system from the coherence (or purity) of one of its subsystems.
That way we utilise a much more general entanglement quantifier, namely the Schmidt number $K$, which is connected to the effective number of modes. 
In order to understand how this can be done, consider a pure state $\ket{\Psi}$ of a bipartite system. Any such state can be decomposed in a certain basis so that \cite{Ekert1995} 
\begin{equation}
\label{eqn:Schmidt:Schmidt_def}
\ket{\Psi} = \sum_{n} \sqrt{\lambda_n} \ket{u_n} \ket{v_n}. 
\end{equation}
The number of non-zero elements $\lambda_n$ required to express the state vector in terms of the Schmidt basis $\{\ket{u_n} \ket{v_n}\}$ is directly related to the lack of separability of the state as well as to the impurity of the subsystems. For infinitely large Hilbert spaces the effective Schmidt number is defined as $K = 1/ \sum_n \left( \lambda_n^2 \right)$ \cite{Grobe1994}. Unlike the Fedorov ratio, the Schmidt number will always give the full amount of entanglement of a system. This is because a Schmidt number greater than one is a direct consequence of the very definition of entanglement (non separability of the state). The Fedorov ratio relies on several assumptions (like Gaussian two-photon amplitudes for instance) in order to quantify EPR like correlations and is, as we have seen, only applicable under certain circumstances. The Schmidt number on the other hand is a more fundamental quantity and does not suffer from any of those limitations. The drawback however is that $K$ is usually not considered an operational measure, meaning it cannot be easily obtained by a direct measurement in many cases. Nevertheless, we were able to implement an experiment which can achieve this. We can apply this Schmidt decomposition to our SPDC state (\ref{eqn:FedRatio:PDCstate}) and obtain
\begin{equation}
\label{eqn:Schmidt:Psi_Schmidt}
\ket{\Psi} = \sum_n \int \mathrm{d} r_s \int \mathrm{d} r_i \phi_n(r_s) \psi_n(r_i) \cre_{r_s} \bcre_{r_i} \ket{0}
\end{equation}
where $\cre_{r_s}$ ($\bcre_{r_i}$) is the creation operators for a signal (idler) photon at position $r_s$ ($r_i$). Hence the Schmidt modes for the SPDC state in terms of the old modes are:
\begin{eqnarray}
\label{eqn:Schmidt:Schmidt_modes}
\hat{A}^{\dagger}_{n} \ket{0}&=& \int \mathrm{d} r_s \phi_n (r_s) \cre_{r_s} \ket{0}\\
\hat{B}^{\dagger}_{n} \ket{0}&=& \int \mathrm{d} r_i \psi_n (r_i) \bcre_{r_i} \ket{0}
\end{eqnarray}
which correspond to $\ket{u_n}$ and $\ket{v_n}$ in equation (\ref{eqn:Schmidt:Schmidt_def}). Since the Schmidt decomposition is orthonormal, 
the new operators also satisfy the commutation relation $\left[ \hat{A}_n, \hat{A}^{\dagger}_m \right] = \delta_{n,m}$. Thus the state now takes the simple form
\begin{equation}
\label{eqn:Schmidt:Psi_Schmidt2}
\ket{\Psi} = \sum_n \sqrt{\lambda_n} \hat{A}^{\dagger}_{n} \hat{B}^{\dagger}_{n} \ket{0}.
\end{equation}
The main idea of the scheme is to obtain the degree of entanglement between signal and idler photons by measuring the coherence (and thus the purity) of one subsystem. To this end let us express the signal field in terms of the Schmidt basis $\hat{E}_s^{(-)} (r) = \sum_n \hat{A}^{\dagger}_n \phi_n(r)$ (and $\hat{E}_s^{(+)} (r) = \sum_n \hat{A}_n \phi_n^{*}(r)$) and consider the first-order correlation function of the signal subsystem 
\begin{equation}
\label{eqn:Schmidt:G1}
\mathrm{G}^{(1)}_s (r,r') = \bra{\Psi} \hat{E}^{(-)} (r) \hat{E}^{(+)} (r') \ket{\Psi} = \sum_n \lambda_n \phi_n (r) \phi^{*}_n (r'). 
\end{equation}

\begin{figure}[<+htpb+>]
\begin{center}
\subfigure[First three Schmidt modes]{
\label{fig:Schmidt:schmidt_modes}
\includegraphics[width=0.4\textwidth]{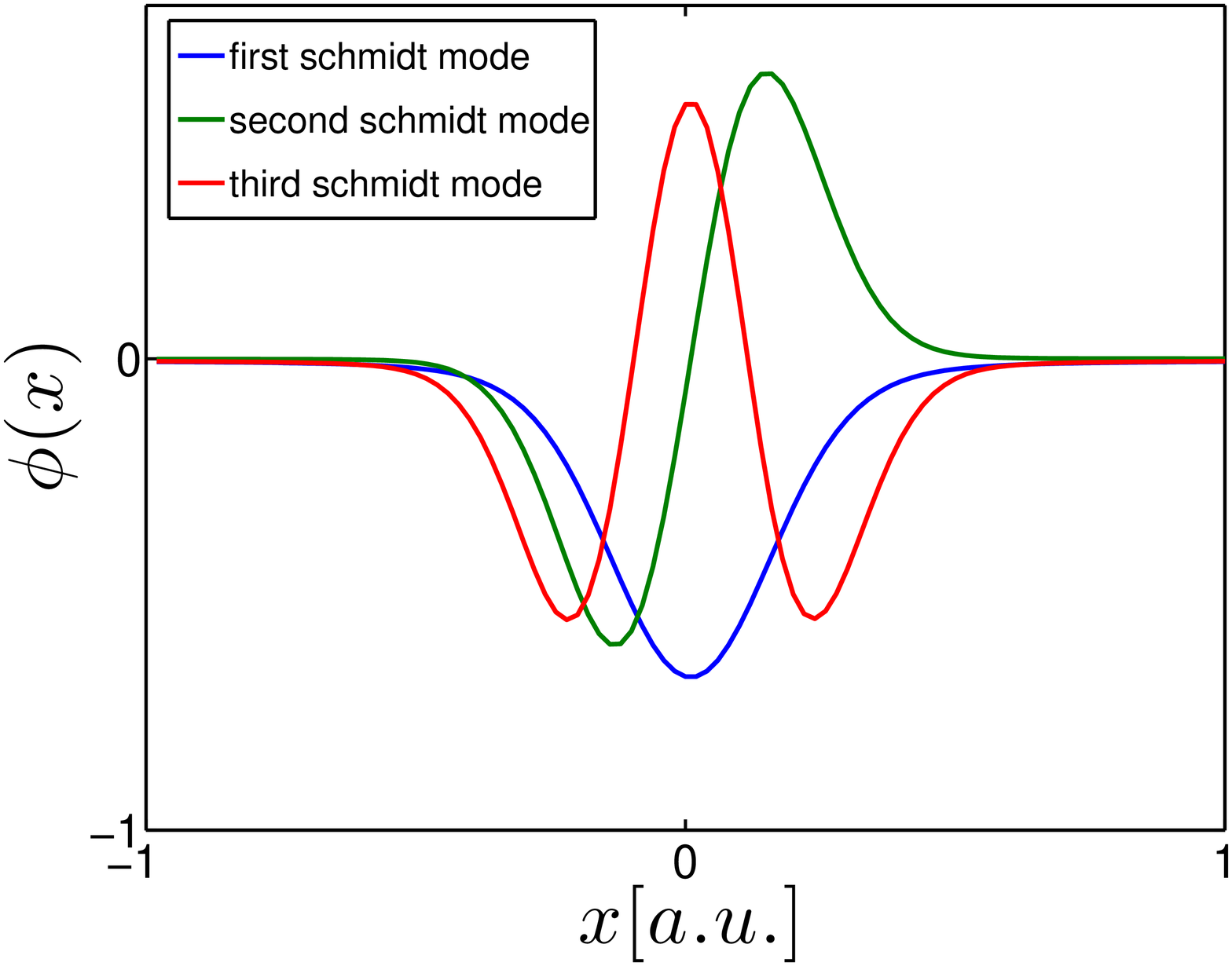}
}
\subfigure[Schmidt coefficients]{
\includegraphics[width=0.4\textwidth]{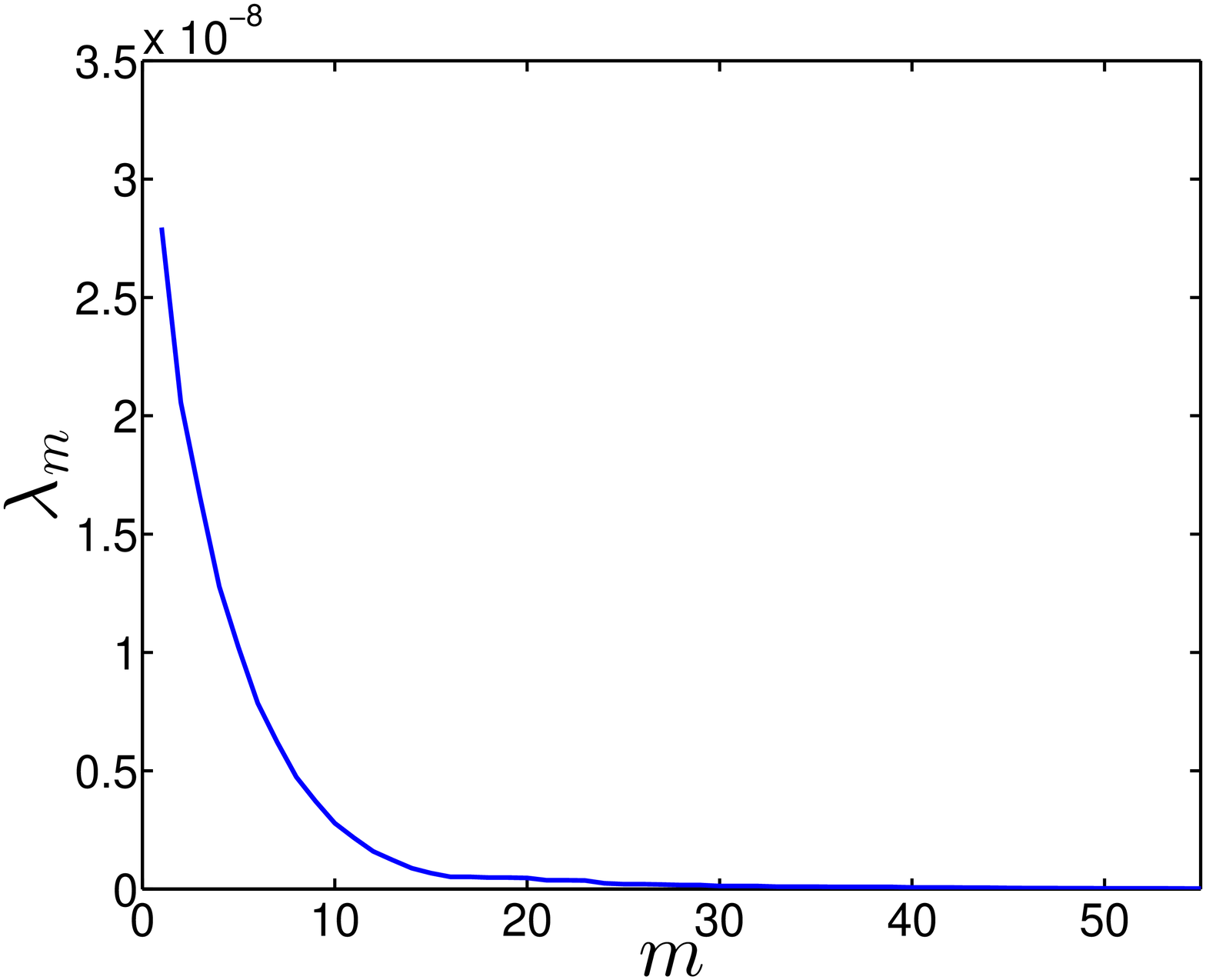}
\label{fig:schmidt_coeff}
}
\end{center}
\caption{Schmidt decomposition of the biphoton state, obtained from a numerical simulation.}
\label{fig:schmidt}
\end{figure}

\begin{figure}[<+htpb+>]
\centering
\subfigure[Two-photon intensity $P_{\cre_s \cre_i}  \left( x_s, x_i \right)$]{
\label{fig:tpaInterferConstr}
\includegraphics[width=0.4\textwidth]{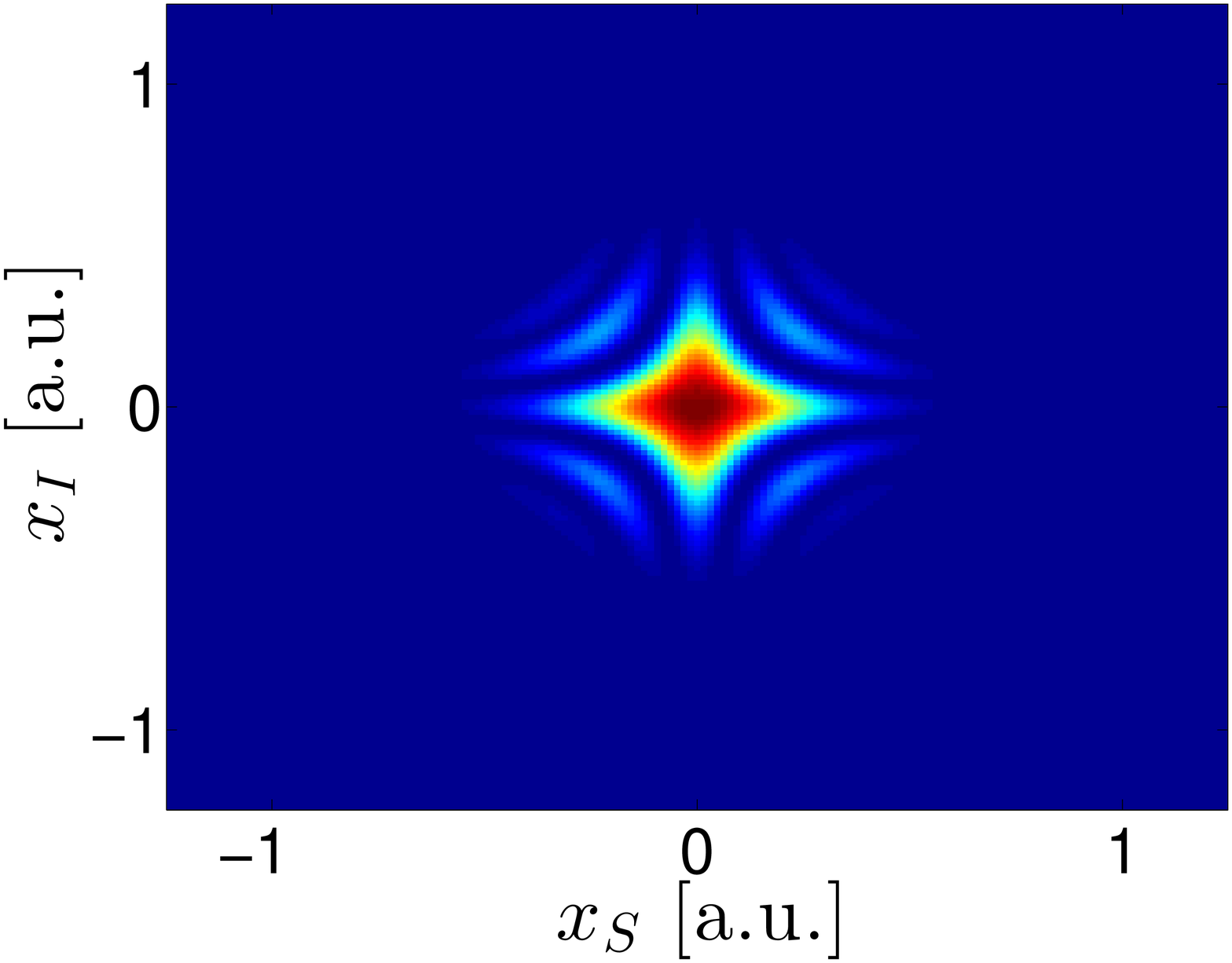}
}\subfigure[Two-photon intensity $P_{\bcre_s \cre_i}  \left( x_s, x_i \right)$ ]{
\label{fig:tpaInterferDestr}
\includegraphics[width=0.4\textwidth]{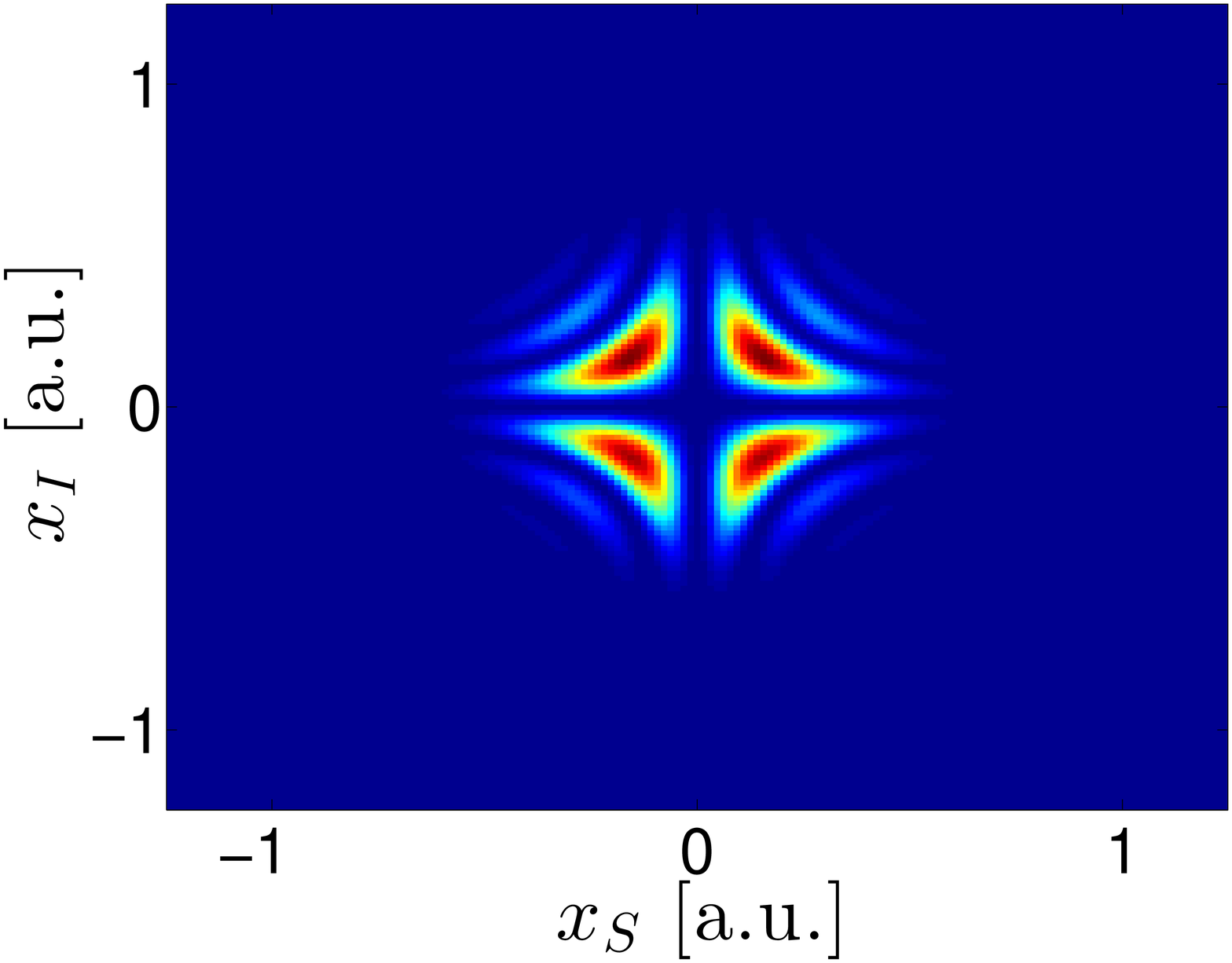}
}
\caption{Simulation of the two-photon intensities ($z=1440$ mm) at both outputs of the interferometer, \ref{fig:tpaInterferConstr} shows constructive interference, \ref{fig:tpaInterferDestr} depicts the complementary destructive interference between the interferometer arms.}
\label{fig:simK}
\end{figure}

\begin{figure}[<+htpb+>]
\begin{center}
\includegraphics[width=0.8\textwidth]{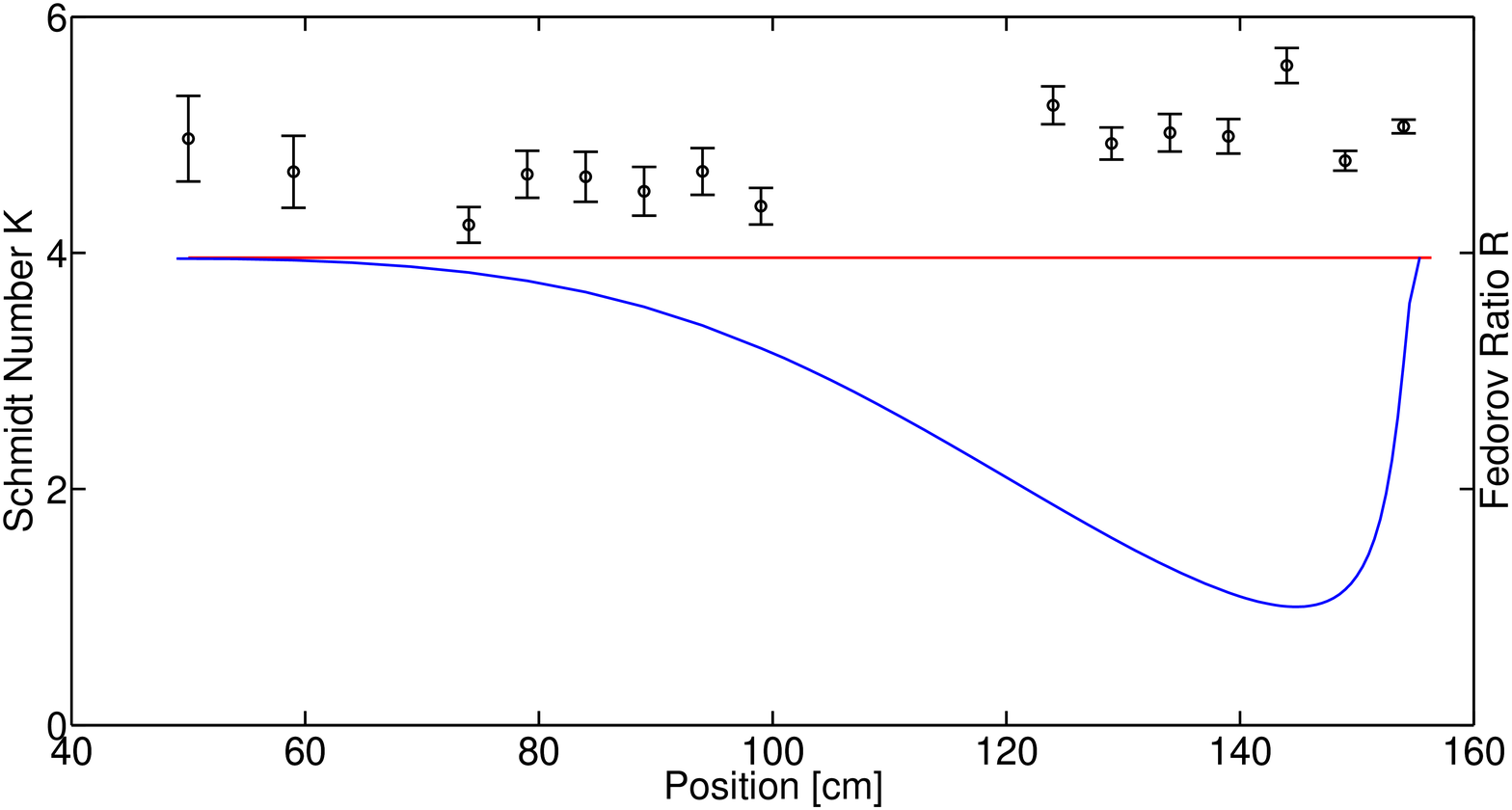}
\end{center}
\caption{Circles are measurements of the Schmidt number, the solid red line is the Schmidt number obtained from a numerical singular value decomposition of the simulated state. For comparison the numerical simulation of the Fedorov ratio is included as well (solid blue line).}
\label{fig:resultsK}
\end{figure}

The effect of the modified Mach-Zehnder interferometer is to overlap the signal with its own spatially inverted copy. Mathematically this can be described as
\begin{equation}
\label{eqn:Schmidt:G1_dove}
\int \mathrm{d} r \mathrm{G}^{(1)}_s \left( r, -r \right) = \sum_n \lambda_n \int \mathrm{d} r \phi_n (r) \phi_n^{(*)} (-r).
\end{equation}
It has been shown that the Schmidt modes for the SPDC biphoton state are given by the Hermite Gaussian polynomials \cite{Fedorov2009,Straupe2011}. Therefore they show a certain symmetry:
\begin{eqnarray}
\label{eqn:Schmidt:symmetry}
\phi_{2m} (r) &=& \phi_{2m} (-r) \\
\phi_{2m+1} (r) &=& -\phi_{2m+1} (-r). 
\end{eqnarray}
In order to illustrate this fact, a plot of the numerically obtained first three Schmidt modes for the SPDC state is given in figure \ref{fig:Schmidt:schmidt_modes}.
Due to these symmetries it is convenient to take odd and even contributions to the sum (\ref{eqn:Schmidt:G1_dove}) into account separately. It thus follows that
\begin{eqnarray}
\label{eqn:SChmidt:G1_dove_2}
\int \mathrm{d}r \mathrm{G}^{(1)}_s \left( r, -r \right) &=& \sum_m \left(  \lambda_{2m} \int \mathrm{d} r  \left| \phi_{2m} (r)\right|^2 - \lambda_{2m+1} \int \mathrm{d} r \left| \phi_{2m+1} (r)\right|^2  \right) \nonumber \\ 
&=& \sum_m \left(  \lambda_{2m} - \lambda_{2m+1} \right).
\end{eqnarray}
Since the eigenvalues of the Schmidt modes decrease exponentially (see figure \ref{fig:schmidt_coeff}) with increasing $m$, \cite{Straupe2011} we can write $\lambda_m = \lambda_0 \alpha^m$ and further use the normalisation condition $\sum_m \lambda_m = 1$ to obtain $\lambda_0 = 1 - \alpha$. So we finally get
\begin{equation}
\label{eqn:Schmidt:G1_dove_3}
\int \mathrm{d}r \mathrm{G}^{(1)}_s \left( r, -r \right) = \lambda_0 \sum_m \left( \alpha^{2m}  - \alpha^{2m+1} \right) = \sum_m \lambda_m^2 = \frac{1}{K}.
\end{equation}
Thus the Schmidt number is inversely proportional to the first-order coherence and therefore to the visibility of interference at the interferometer output which can be measured in the laboratory.
This result is in perfect agreement with \cite{Chan2007} where it was suggested to measure   
\begin{equation}
\label{eqn:Schmidt:Kvis}
K = \frac{(P_+ + P_-)}{(P_+ - P_-)},
\end{equation} 
where $P_+$ and $P_-$ are the conditional count rates 
\begin{eqnarray}
\label{eqn:Ppm}
P_+ = \int \int \mathrm{d} x_s \mathrm{d} x_i P_{\hat{c}^{\dagger} \bcre} \left( x_s, x_i \right)\\
P_- = \int \int \mathrm{d} x_s \mathrm{d} x_i P_{\hat{d}^{\dagger} \bcre} \left( x_s, x_i \right). 
\end{eqnarray}
Here $P_{\hat{c}^{\dagger} \bcre} \left( x_s, x_i \right)$ denotes the probability to observe the signal photon at the output $\hat{c}^{\dagger}$ (constructive port) of the interferometer at position $x_s$ and the idler photon at position $x_i$ in the idler mode $\cre_i$. Analogously $P_{\hat{d}^{\dagger} \bcre} \left( x_s, x_i \right)$ describes the joint detection probability between the other interferometer output $\hat{d}^{\dagger}$ (destructive port) and idler $\cre_i$ at $\left( x_s, x_i \right)$.\\
Figure \ref{fig:simK} shows a numerical simulation of the two probability distributions at both interferometer outputs. According to equation (\ref{eqn:Schmidt:Kvis}), the Schmidt number is given by the inverse visibility of those coincidence probabilities integrated over $x_s$ and $x_i$. 
We would like to remark that, using this method, in general measurements on one subsystem is sufficient to determine the degree of entanglement. Thus coincidence measurements as proposed by \cite{Chan2007} are not strictly necessary. In a real experiment however it is absolutely crucial to reduce the amount of noise as much as possible since naturally, the absence of visibility is very hard to detect. Accordingly it is much more convenient to perform coincidence measurements. Unlike other schemes, our setup enables one to measure the full amount of entanglement in any arbitrary position in a single measurement without the need to perform measurements at specific positions, such as the near and far field \cite{Pires2009}.\\ 
Instead of using two detectors as indicated in figure \ref{fig:setup} we measured the visibility in coincidences by scanning the phase of the interferometer with the help of a piezo attached to one of the mirrors. This time we position two $200 \mu m$ slits oriented along the $x$ direction in both signal and idler arms. This is equivalent to an integration along $x_s$ and $x_i$ in one single measurement. Hence we are able to determine the Schmidt number simply by moving the piezo from the position of maximum count rate to minimum count rate in one output of the interferometer and record the visibility. Both the results of the measurements as well as the numerical predictions are depicted in figure \ref{fig:resultsK}.       
It is apparent that the measurement points almost exclusively lie above the numerical calculation, which is not surprising because the entanglement we are trying to determine actually manifests itself in the lack of visibility. Accordingly any experimental imperfection always will lead to an overestimation of entanglement and consequently a slightly increased Schmidt number. In figure (\ref{fig:resultsK}) the numerical Fedorov ratio does not agree with the Schmidt number even in near and far field. The reason for this is that the equality between the two measures only holds in the double Gaussian approximation \cite{Fedorov2006} (where it can even be proven analytically). In reality however, the shape of the two-photon amplitude is governed by the product of a Gaussian and a $\mathrm{sinc}$ function, which is taken into account by our numerical model. \\

\section{Conclusion}
\label{conclusion}
As mentioned before, the reason for the breakdown of $R$ is that upon propagation in free space the two-photon amplitude acquires a quadratic phase \cite{Chan2007}. 
Henceforth, this leads to a shift of entanglement from the modulus to the phase of the quantum state at a certain position. An interesting implication arises from the analogy, that this quadratic phase is mathematically identical to the one acquired by the two-photon amplitude due to the group velocity dispersion in an optical fibre \cite{Valencia2002}. Thus it should be possible to obtain similar results in the time-frequency entanglement of spontaneous parametric downconversion photons using a fibre as dispersive medium. Such effects could be an issue in QKD experiments where two-photon states are sent through very long fibres.\\
In conclusion we have experimentally demonstrated that the Fedorov ratio cannot correctly assess the entanglement in the transverse momentum of the biphoton state emitted by parametric down-conversion at arbitrary distances from the source. Responsible for this is the fact that the entanglement can partially (and under special circumstances even fully) reside in the phase of the quantum state. Naturally the Fedorov ratio, being a measure of intensity correlation, is therefore not sensitive to the full entanglement of the system. We present experimental and numerical results showing the migration of entanglement from the modulus to the phase of the two-photon amplitude upon free space propagation. Furthermore we implement a measurement technique, which overcomes the shortcoming of the previous approach by taking into account both phase and modulus of the state at the same time. It is shown that this scheme allows a direct measurement of the Schmidt number and thus by definition of the degree of entanglement of the quantum state in question. We believe that this kind of experiment provides useful insight into the nature of the non classical correlations of PDC which might be generalised to other systems \cite{Fedorov2005} in the future.
\ack
The authors would like to thank Farid Khalili for many helpful discussions. This work is supported by the  ERA-Net.RUS (project Nanoquint)

\section*{References}
\bibliographystyle{unsrt}



\begin{thebibliography}{10}

\bibitem{Law2004}
Law K and Eberly J 2004 \textit{Phys. Rev. Lett.} \textbf{92} 127903

\bibitem{Strekalov1995}
Strekalov D, Sergienko A, Klyshko D, and Shih Y 1995 \textit{Phys. Rev. Lett.} \textbf{74} 3600--3603

\bibitem{Monken1998}
Monken C, Souto Ribeiro P and P\'adua S 1998 \textit{Phys. Rev. A} \textbf{57} 3123--3126

\bibitem{Menzel2012}
Menzel R, Puhlmann D, Heuer A, and Schleich W 2012 \textit{Proceedings of the National Academy of Sciences} \textbf{109}(24) 9314--9319

\bibitem{Lloyd1999}
Lloyd S and Braunstein S 1999 \textit{Phys. Rev. Lett.} \textbf{82} 1784--1787

\bibitem{Chiuri2012}
Chiuri A, Greganti C, Paternostro M, Vallone G, Mataloni P 2012 \textit{Phys. Rev. Lett.} \textbf{109} 173604

\bibitem{Howell2012}
Dixon P, Howland G, Schneeloch J, and Howell J 2012 \textit{Phys. Rev. Lett.} \textbf{108} 143603

\bibitem{Kang2012}
Kang Y, Ko J, Lee S, Choi S, Kim B and Park H 2012 \textit{Phys. Rev. Lett.} \textbf{109} 020502

\bibitem{Salakhutdinov2012}
Salakhutdinov V, Eliel E, and L\"offler W 2012 \textit{Phys. Rev. Lett.} \textbf{108} 17360

\bibitem{Braverman2012}
Braverman B  and Simon C 2013 \textit{Phys. Rev. Lett.} \textbf{110} 060406

\bibitem{EPR}
Einstein A, Podolsky B, and Rosen N 1935 \textit{Phys. Rev.} \textbf{47} 777--780

\bibitem{Howell2004}
Howell J, Bennink R, Bentley S, and Boyd R 2004 \textit{Phys. Rev. Lett.} \textbf{92} 210403

\bibitem{Leach2012}
Leach J, Warburton R, Ireland D, Izdebski F, Barnett S, Yao A, Buller G, and Padgett M 2012 \textit{Phys. Rev. A} \textbf{85} 013827

\bibitem{Fedorov2004}
Fedorov M, Efremov M, Kazakov A, Chan K, Law C, and Eberly J 2004 \textit{Phys. Rev. A} \textbf{69} 052117

\bibitem{Brida2009}
Brida G, Caricato V, Fedorov M, Genovese M, Gramegna M, and Kulik S 2009 \textit{EPL} \textbf{87} 64003

\bibitem{Chan2007}
Chan K, Torres J, and Eberly J 2007 \textit{Phys. Rev. A} \textbf{75} 050101

\bibitem{Ekert1995}
Ekert A and Knight P 1995 \textit{Am. J. Phys.}, \textbf{63} 415

\bibitem{Klyshko1988}
Klyshko D 1998 \textit{Photons and Nonlinear Optics} Gordon and Breach

\bibitem{Tasca2008}
Tasca D, Walborn S, Souto~Ribeiro P, and Toscano F 2008 \textit{Phys. Rev. A} \textbf{78} 010304

\bibitem{Tasca2009}
Tasca D, Walborn A, Souto~Ribeiro P , Toscano F, and Pellat-Finet P 2009 \textit{Phys. Rev. A} \textbf{79} 033801

\bibitem{Sych2009}
Sych D and Leuchs G 2009 \textit{NJP} \textbf{11} 013006

\bibitem{Leuchs2009}
Leuchs G, Ruifang D, and Sych D 2009 \textit{NJP} \textbf{11} 113040

\bibitem{Grobe1994}
Grobe R, Rz\k{a}\.zewski K, and Eberly J 1994 \textit{J. Phys. B} \textbf{27} L503--8

\bibitem{Pires2009}
Di~Lorenzo~Pires H, Monken C, and van Exter M 2009 \textit{Phys. Rev. A} \textbf{80} 022307

\bibitem{Valencia2002}
Valencia A, Chekhova M, Trifonov A, and Shih Y 2002 \textit{Phys. Rev. Lett.} \textbf{88} 183601

\bibitem{Fedorov2009}
Fedorov M, Mikhailova Y, and Volkov A 2009 \textit{J. Phys. B} \textbf{42} 175503

\bibitem{Straupe2011}
Straupe S, Ivanov D, Kalinkin A, Bobrov I, and Kulik S 2011 \textit{Phys. Rev. A} \textbf{83} 060302

\bibitem{Fedorov2006}
Fedorov M, Efremov M, Volkov A, and Eberly J 2006 \textit{J. Phys. B} \textbf{39} S467

\bibitem{Fedorov2005}
Fedorov M, Efremov M, Kazakov A, Chan K, Law K, and Eberly J 2005 \textit{Phys. Rev. A} \textbf{72} 032110

\end{thebibliography}


\end{document}